\documentclass[12pt,twoside]{article}

\usepackage{amsmath,amssymb,amsbsy,amsfonts,amsthm,latexsym,amsopn,amstext,amsxtra,euscript,amscd}

\date{}

\begin{document}


\centerline{}

\centerline{}

\centerline {\Large{\bf New Variant of ElGamal Signature Scheme}}

\centerline{}



\centerline{\bf {Omar Khadir}}

\centerline{}

\centerline{ Department of Mathematics,}

\centerline{ Faculty of Science and Technology,}

\centerline{ University of Hassan II-Mohammedia, Morocco.}


\centerline{khadir@hotmail.com}

\centerline{}







\newtheorem{Theorem}{\quad Theorem}[section]

\newtheorem{Definition}[Theorem]{\quad Definition}

\newtheorem{Corollary}[Theorem]{\quad Corollary}

\newtheorem{Lemma}[Theorem]{\quad Lemma}

\newtheorem{Example}[Theorem]{\quad Example}

\newtheorem{Proposition}[Theorem]{\quad Proposition}

\newtheorem{Remark}[Theorem]{\quad Remark}

\begin{abstract} In this paper, a  new variant of ElGamal signature scheme
is presented and its security  analyzed. We also give, for its
theoretical interest,  a general form of the signature equation.

\end{abstract}




\section{Introduction} Since the invention of the public key cryptography
in the late 1970s [2, 13, 12], several new subjects related to the
data security as identification, authentication, zero-knowledge
proof and secret sharing were explored. But among all these
issues, and perhaps the most important, is how to build secure
digital signature systems. During more than three decades, the
topic, probably due to its fundamental and practical role in
electronic funds transfer, was intensively investigated
[10, 15, 14, 4, 1, 11, 9].\\
There is only one principle on which rest the digital signature
algorithms.  To sign a message $m$, Alice with the help of her
private key, must answer a question asked by Bob, the verifier.
The question is naturally a function of $m$. Nobody other than
Alice is able to forge her signature and give
the right answer, even the asker himself.\\
In most digital signature schemes, the considered question is a
difficult mathematical equation depending of $m$ as a parameter.
Only Alice, because she possesses a private key, is able to solve
it. In this protocol, we are not necessary concerned by the
transmitted data security. Indeed,  Bob and Alice can publish
respectively the equation and the solution in two
protected and separated personal servers.\\
In 1985, ElGamal [3], inspired by the Diffie-Hellman
 ingenious ideas on new directions in  cryptography [2], was one of
the firsts  to propose a practical signature scheme. Used
properly, this signature system has never been broken. He built it
on a simple equation with two unknown variables. The hardness of
this equation relies on the discrete logarithm problem [7, p.103].
In general, from a  public key cryptosystem, one can derive a
signature scheme. Curiously, in his paper [3], ElGamal did not
exploit this possibility and it is  still unclear how he found his
signature equation. This fact has encouraged many researchers to
look for equations having properties similar to
those of ElGamal. See, for instance, [14, 4, 5]. \\
Some practical signature protocols as Schnorr method [14] and the
digital signature algorithm DSA [8] are directly derived from
ElGamal scheme. \\
Permanently, ElGamal signature scheme is facing attacks more and
more sophisticated. If the system is completely broken,
alternative protocols, previously designed, prepared and tested,
would be useful. In this work we present a new variant of the
ElGamal signature method and analyze its security. Furthermore, we
give, just for its theoretical interest, a general form of our
signature equation.\\
The paper is organized as follows. In section 2, we review the
basic ElGamal signature algorithm and recall the main known
attacks. Our new variant and a theoretical generalization are
presented in  section 3. We conclude in section~4. \\
In the sequel, we will adopt ElGamal paper notations [3].
$\mathbb{Z}$, $\mathbb{N}$ are respectively the sets of  integers
and non-negative integers. For every positive integer $n$, we
denote by $\mathbb{Z}_n$ the finite ring of modular integers and
by $\mathbb{Z}_n^*$ the multiplicative group of its invertible
elements. Let $a,b,c$  be three integers. The great common divisor
of $a$ and $b$ is denoted by $gcd(a,b)$. We write $a\equiv b$
$[c]$ \ if $c$ divides the difference $a-b$, and $a=b\ mod\ c$ if
$a$ is
the remainder in the division of $b$ by $c$.\\
We start by describing the  original ElGamal signature
 scheme.


\section{ ElGamal Original  Signature Scheme}
We recall in this section the basic ElGamal protocol in three
steps, followed by the most known attacks.

\vspace{0.5cm} \noindent {\bf 2.1. ElGamal Algorithm }

\vspace{0.2cm} \noindent {\bf 1.} Alice begins by choosing three
numbers :

\hspace{1cm} - $p$, a large prime integer.

\hspace{1cm} - $\alpha$, a primitive root [7, p.69] of the finite
multiplicative group $\mathbb Z_p^*$.

\hspace{1cm} - $x$, a random element in $\{1,2,\ldots,p-1\}$.

\noindent She computes  $y=\alpha^x\ mod\ p$. We consider then
that : $(p,\alpha,y)$ is Alice public key and $x$ her private key.

\vspace{0.2cm} \noindent
 {\bf 2.} Assume that Alice wants to sign the message  $m<p$. She must
 solve the congruence
\begin{equation}
 \alpha^m \equiv y^r\,r^s\ [p]
\end{equation}
where $r$ and $s$ are two unknown variables. \\
Alice fixes arbitrary $r$ to be $r=\alpha^k\ mod\ p$, where $k$ is
chosen randomly and invertible modulo $p-1$. She has exactly
$\varphi(p-1)$ possibilities for $k$, where $\varphi$ est the
phi-Euler function [7, p.65]. Equation (1) is then equivalent to :
\begin{equation}
 m\equiv x \, r+k\, s \ [p-1]
\end{equation}
As Alice possesses the secret key $x$, and as the integer $k$ is
invertible modulo $p-1$, she  computes the second unknown variable
$s$ by : $\displaystyle s\equiv \frac{m-x\, r}{k}\ [p-1]$

\vspace{0.2cm} \noindent
 {\bf 3.} Bob can verify the signature by checking  that congruence
 (1) is valid.

\vspace{0.2cm} \noindent  Keys generation  problem must be taken
into account. There exist essentially probabilistic algorithms for
generating prime integers. In a recent previous
work [6], we obtained experimental results on the subject.\\
Now, we recall the main known attacks.

\vspace{0.5cm}\noindent {\bf 2.2. Main attacks }

\vspace{0.2cm}\noindent The first attack was mentioned by ElGamal
himself [3]. It is not recommended to sign two different messages
with the same secret exponent. As the complete justification of
this attack does not figure in the ElGamal paper, we reproduce
here the proof from [16, p. 291] which seems to us, less
restrictive than that in [7, p.455].

\begin{Proposition}
If Alice signs more than one message with the same secret
exponent, then her system can be totally broken. \end{Proposition}
 \proof  Let $(m_1,r,s_1)$ and $(m_2,r,s_2)$ be the  signatures
of the two messages $m_1$ and $m_2$ with the same secret exponent
$k$. Due to relation (2), we retrieve Alice secret key $x$ if we
 find the value of the parameter  $k$ provided that $r$ is invertible modulo $p-1$.\\
We have $m_1\equiv x\, r + k\, s_1\ [p-1]$ and $m_2\equiv x\, r +
k\, s_2\ [p-1]$, so :
\begin{equation}
m_1-m_2\equiv k\,(s_1-s_2)\ [p-1]
\end{equation}
If we put $gcd(s_1-s_2,p-1)=d$, there exist two integers $S$ and
$P$ such that $s_1-s_2=d\, S$, $p-1=d\, P$ and $gcd(S,P)=1$. Thus
 relation (3) becomes :\\
 $ m_1-m_2= k\,(s_1-s_2)+K\, (p-1)=k\,d\,S+K\,d\,P,\  K\in \mathbb
 Z$. With  $M=k\,S+K\, P$, we obtain $M\equiv k\, S\ [P]$. As
 $S$ is invertible modulo $P$, we have
\begin{equation}
 k=M\, S^{-1}+K\, P
\end{equation}
Since $k<p-1$ and $p-1=d\, P$,  we  deduce  that $K<d$. By
 equality (4), we can test every value of $K$ and check if
 $r\equiv \alpha^k\ [p]$. We find $K$ if $d$ is not too large.\\
 \qed

\vspace{0.5cm}\noindent In 1996, Bleichenbacher [1] has discovered
an important fact : when some parameters are smooth [16, p.197],
it is possible to forge ElGamal signature without solving the
discrete logarithm problem. We present here a slightly
 modified version of his result.
\begin{Proposition}
Let $(p,\alpha,y)$ be Alice public key. Suppose that $\beta<p$ is
a positive integer for which one can efficiently compute $t\in
\mathbb N$ such that
$\alpha\equiv \beta^t\ [p]$.\\
 If $\displaystyle
\frac{p-1}{gcd(p-1,\beta)}$ is smooth, then an Alice adversary
will be  able to forge her signature for any given message $M$.
\end{Proposition}
\proof Let $D=gcd(p-1,\beta)$ and $\beta=\lambda\, D, \
\lambda\in\mathbb N^*$. We denote by $H$ the subgroup of $\mathbb
Z^*$ generated by $\alpha^D\ mod\ p$. Since $y^D\equiv
(\alpha^x)^D\equiv (\alpha^D)^x\ [p]$, we have $y^D\in H$. From a
well known result, as the order $(p-1)/D$ of H is smooth, the
discrete logarithm problem is computationally feasible
: one can efficiently find  $z_0\in \mathbb N$ such that $y^D\equiv (\alpha^D)^{z_0}\ [p]$.\\
Let $M$ a message to be signed and $m=h(M)\ mod\ p$ where $h$ is a
public hash function. Alice adversary sets $r=\beta$. ElGamal
signature equation (1) becomes~:
$$\beta^{t\, m}\equiv y^\beta\, \beta^s\equiv y^{\lambda\, D}\, \beta^s\equiv
(\alpha^D)^{z_0\, \lambda}\, \beta^s\equiv \beta^{\lambda\, t\,
z_0\, D}\, \beta^s \ [p]$$
 \noindent Hence $s\equiv t\, (m-\beta \, z_0)\ [p-1]$, and then
 the couple  $(r,s)$ is a valid signature of the message $M$,
 which achieves the proof.\\
Observe that it is not so surprising to choose $r=\beta$  or
$r=\beta^i\ mod \ p, i\in\mathbb N$,  since $\beta^t\equiv \alpha
\ [p]$
implies that $\beta$ is an other generator of  $\mathbb Z_n^*$.\\
\qed

\vspace{1cm}\noindent
 Next section presents our main contribution.

\section{New  Variant and  Theoretical  Generalization}
\vspace{0.3cm} In this section, we suggest a new variant of
ElGamal signature scheme based on an equation with three unknown
variables. The method does not need the computation of the secret
exponent inverse and so avoids the use of the extended Euclidean
algorithm. Technical report [4], although it collected several
signature equations, did not study  the case we propose here.

\vspace{0.5cm} \noindent {\bf 3.1. Our protocol}

\vspace{0.2cm} \noindent We suppose first that  $h$ is a public
secure hash function. We can take $h$ equal to the secure
hash algorithm SHA1 [7, Chap.9] and [16, Chap.5].\\
 {\bf 1.} Alice begins by choosing her
public key $(p,\alpha, y)$, where $p$ is a large prime integer,
$\alpha$ is a primitive element of the finite multiplicative group
$\mathbb Z_p^*$ and $y=\alpha^x\ mod\ p$. Element $x$, which is a
random  integer in $\{1,2,3,\ldots,p-1\}$, is Alice private key.

\vspace{0.2cm} \noindent
 {\bf 2.} Assume that Alice wants to sign the message  $M<p$. She must
 solve the congruence
\begin{equation}
 \alpha^t \equiv y^r\,r^s\, s^m \  [p]
\end{equation}
where $r,s$ and $t$ are three  unknown variables and $m=h(M)\ mod\ p$. \\
Alice fixes arbitrary $r$ to be $r=\alpha^k\ mod \ p$, and $s$ to
be
$s=\alpha^l\ mod \ p$, where $k,l$ are chosen randomly in $\{1,2,\ldots,p-1\}$.\\
 Equation (5) is then
equivalent to :
\begin{equation}
  t\equiv r\,x+ks+l\, m\
[p-1].
\end{equation}
As Alice detains the secret key $x$ and knows the values of
$r,s,k,l,m$, she is able to compute the  third unknown variable
$t$.

\vspace{0.2cm} \noindent
 {\bf 3.} Bob can verify the signature by checking  that congruence
 (5) holds.\\

Our scheme has the advantage that  it does not  need the use of
the  extended Euclidean algorithm for  computing $k^{-1}$ modulo
$p-1$. May be this can be an answer  to problems evoked in [9,
subsection 1.3].

 \vspace{0.2cm} \noindent
To illustrate the technique, we give the following small example.

\vspace{0.2cm} \noindent
 \begin{Example}
 Let $(p,\alpha,y)$ be Alice public key where : $p=509$,
 $\alpha=2$ and $y=482$. We emphasize that we are not sure if using a short value
 of $\alpha$ does not
 weaken  the system. The private key is $x=281$. Suppose that Alice
wants to produce a
 signature  for the message $M$ for which  $m\equiv h(M)\equiv 432\ [508]$ with the
 two random exponents $k=208$ and $l=386$. She
 computes  $r\equiv \alpha^k\equiv 2^{208}\equiv 332\ [p]$,
 $s\equiv \alpha^l\equiv 2^{386}\equiv 39\ [p]$ and $t\equiv
 r\,x+k\,s+l\,m\equiv 440\ [p-1]$. Bob or anyone can verify the
 relation $\alpha^t \equiv y^r\,r^s\, s^m \  [p]$. Indeed, we find
 that $\alpha^t\equiv 436\ [p]$ and $y^r\,r^s\, s^m \equiv 436 \
 [p]$. Notice here that $k$ and $l$ are even integers unlike in ElGamal protocol where
 the exponent $k$ is always odd since it must be relatively prime with
 $p-1$.
 \end{Example}

\vspace{0.3cm} \noindent
 {\bf 3.2. Security analysis}

 \vspace{0.2cm} \noindent
 Suppose that Oscar is an Alice adversary. Let us discuss some
 possible and realistic attacks.

\vspace{0.3cm}\noindent {\bf Attack 1 : } Knowing all signature
parameters for a particular message $M$, Oscar tries to find Alice
secret key $x$.\\
Equation (5) is equivalent to $\alpha^t\equiv \alpha^{x\,
r}\,r^s\, s^m\ [p]$, so ${\alpha^r}^x\equiv \alpha^t\, r^{-s}\,
s^{-m} \ [p]$. Therefore, Oscar is confronted to the hard discrete
logarithm problem.\\
If Oscar prefers to work with relation (6), he needs to know $k$
and $l$. Their computation conducts to the discrete logarithm
problem.

\vspace{0.3cm}\noindent {\bf Attack 2 : } Oscar tries to forge
Alice signature for a message $M$, by first, fixing arbitrary two
unknown variables and looking for the third parameter.

\vspace{0.2cm}\noindent {\bf (1)}  Suppose for example that Oscar
has fixed $r,s$, and tries to solve equation (5) in the variable
$t$. But here again, he will be confronted to the discrete
logarithm problem.

\vspace{0.2cm}\noindent {\bf (2)}  Assume that Oscar has fixed $r$
and $t$. We have from relation (5): $r^s \, s^m\equiv \alpha^t\,
y^{-r} \ [p]$; and there is no known way to solve this equation.

\vspace{0.2cm}\noindent {\bf (3)}  Assume now that Oscar has fixed
$s$ and $t$. We have from relation (5)~: $y^r \, r^s\equiv
\alpha^t\, s^{-m} \ [p]$; and this equation is similar to the last
case, so it is intractable.

\vspace{0.3cm}\noindent {\bf Attack 3 : } Let us admit that Oscar
has collected $n$ valid signatures for messages $M_i$,
$i\in\{1,2,3,\ldots,n\}$ and $n\in \mathbb N$. He will obtain a
system of $n$ modular equations :
$$(S)\left\{%
\begin{array}{c}
  t_1\equiv x\, r_1+k_1\,s_1+l_1\,m_1\ [p-1]  \\
  t_2\equiv x\, r_2+k_2\,s_2+l_2\,m_2\ [p-1]  \\
  \vdots \ \ \ \vdots \ \ \  \vdots\\
  t_n\equiv x\, r_n+k_n\,s_n+l_n\,m_n\ [p-1]  \\
\end{array}%
\right.$$ Where  $\forall i\in\{1,2,3,\ldots,n\},$ $r_i\equiv
\alpha^{k_i}\ [p],\ s_i\equiv \alpha^{l_i}\ [p]$ et $m_i\equiv
h(M_i)\ [p]$\\
Since system (S) contains $2n+1$ unknown variables $x,r_i,s_i,\
i\in\{1,2,3,\ldots,n\}$, Oscar can find several valid solutions.
However, as $x$ is Alice secret key, it has a unique possibility
and therefore Oscar will never be sure what value of $x$ is the
correct one. Consequently, this attack is to be rejected.\\
Next result is similar to that exists in ElGamal scheme.

\begin{Proposition}
If no hash function is used, then Oscar can forge existentially
Alice signature. \end{Proposition}
 \proof  Assume that Alice products the parameters  $(r,s,t)$ as a
 signature for the message $M$. So $\alpha^t \equiv y^r\,r^s\, s^m \
 [p]$. Let $k,k',l,l'\in\mathbb N$ be four arbitrary integers with
 $gcd(l',p-1)=1$. If Oscar chooses $r\equiv \alpha^k\, y^{k'}\ [p]$
and $s\equiv \alpha^l\, y^{l'}\ [p]$, he would obtain :
\begin{equation}
\alpha^t \equiv y^r\,(\alpha^{k\, s}\, y^{k'\, s})\, (\alpha^{l\,
m}\, y^{l'\, m}) \
 [p].
\end{equation}
Relation (7) holds if
$\left\{%
\begin{array}{c}
  t-k\,s -l\, m \equiv 0\ [p-1]\ \ \   (7.1)\\
  t-k'\,s -l'\, m\equiv 0  [p-1]\ \ \ (7.2)\\
\end{array}%
\right.$\\
 Oscar computes $m$ from equality (7.2) : $\displaystyle m\equiv
 \frac{r+k'\,s}{l'}\ [p-1]$; and from (7.1) he has $\displaystyle t\equiv k\,s+ \frac{l\, (r+k'\,s)}{l'}\
 [p-1]$.
 Thus $(r,st)$ is a valid signature for the message $m$.\\
\qed

\begin{Remark}
Alice can sign two messages with the same couple of secret
exponents. Indeed, let $(r,s,t_1)$ and $(r,s,t_2)$ be the
signatures of the two different messages $M_1$ and $M_2$
associated to the secret exponents $(k,l)$. We have $\left\{%
\begin{array}{c}
  t_1\equiv x\,r+k\, s+l\,m_1\ [p-1] \\
    t_2\equiv x\,r+k\, s+l\, m_2\ [p-1] \\
\end{array}%
\right.$ \\
where $m_1\equiv h(M_1)\ [p-1]$ et $m_2\equiv h(M_2)\ [p-1]$.\\
We can follow the method used in the proof of Proposition 1 and
find the value of $l$, but it seems that it is not an easy task to
retrieve secret parameters $k$ and~$x$. \end{Remark}

\vspace{0.5cm}\noindent  {\bf 3.3. Complexity of our method :}

\vspace{0.3cm}\noindent As in [5], let $T_{exp},\  T_{mult},\ T_h,
$ be respectively the time to perform a modular exponentiation, a
modular multiplication and  hash function computation of a message
$M$. We ignore the time required for modular additions,
substractions, comparisons and make the conversion $T_{exp}=240\,
T_{mult}$.\\
 The signer Alice needs to perform two modular
exponentiations, three modular multiplications and one hash
function computation. So the global required time is :
$T_1=2\,T_{exp}+ 3\, T_{mult}+T_h=483\,T_{mult}+T_h$. \\
The verifier Bob needs to perform four modular exponentiations,
two modular  multiplications and one hash function computation.
 So the global required time is : $T_2=4\, T_{exp}+ 2\, T_{mut}+T_h=962\,T_{mult}+T_h$. \\
The cost of communication, without $M$, is $6\, |p|$, since to
sign, Alice transmits $(p,\alpha,y)$ and $(r,s,t)$. $|p|$ denotes
the bit-length of the integer $p$.\\
Observe that the complexity of our method is not too high
relatively to that of ElGamal scheme or to that in [5].


\vspace{0.5cm} \noindent {\bf 3.4. Theoretical generalization}

\vspace{0.3cm}\noindent Let $h$ be a public secure hash function.

 \vspace{0.3cm}\noindent  {\bf 1.} Alice
begins by choosing her public key $(p,\alpha, y)$, where $p$ is a
large prime integer, $\alpha$ is a primitive element of the finite
multiplicative group $\mathbb Z_p^*$ and $y=\alpha^x$, $x$ is a
random integer in $\{1,2,3,\ldots,p-1\}$. $x$  is the Alice
private key.

\vspace{0.2cm} \noindent
 {\bf 2.} Assume that Alice wants to sign the message  $m<p$. She must
 solve the congruence
\begin{equation}
 \alpha^t \equiv y^{r_1}\,r_1^{r_2}\, r_2^{r_3}\ldots r_{n-1}^{r_n}\,r_n^m\  [p]
\end{equation}
where $r_1,r_2,\ldots, r_n, t$ are $n+1$  unknown variables. \\
Alice fixes arbitrary $r_1$ to be $r_1=\alpha^{k_1}$, $r_2$ to be
$r_2=\alpha^{k_2}$,..., and $r_n$ to be $r_n=\alpha^{k_n}$,
where $k_1,k_2,\ldots ,k_n$ are chosen randomly.\\
Equation (8) is then equivalent to :
\begin{equation}
  t\equiv x\,r_1+k_1\,r_2+ \ldots+k_{n-1}\,r_n+k_n\, m\
[p-1].
\end{equation}
As Alice detains the secret key $x$ and knows the values
$r_i,k_j,m, \ i\in \{1,2,\ldots, n\}$, she is able to compute the
$(n+1)th$  unknown variable $t$.

\vspace{0.2cm} \noindent
 {\bf 3.} Bob can check that verification condition (8) is valid.

\begin{Remark}
Let $\overrightarrow{u}=(x,k_1,k_2,\ldots, k_n)$ be Alice  secret
keys vector and $\overrightarrow{v}=(r_1,r_2,\ldots,r_n,m)$ the
signature parameters vector. If
$\overrightarrow{u}.\overrightarrow{v}$ denotes the scalar
product, then the last signature parameter $t$ can be obtained
from the modular equation  $t\equiv
\overrightarrow{u}.\overrightarrow{v}\ [p-1]$, which is an
immediate consequence of relation (9). \end{Remark}

\section{Conclusion}
In this work, we described a new variant
 of ElGamal signature scheme and analyzed  its security.
 Our method relies on an ElGamal similar equation with three unknown variables and it
  avoids the use of the extended Euclidean algorithm. We also gave
 a generalization for its theoretical interest.\\
For the future, one may try to see how to improve our
 new variant. One idea is to replace the modular group $\mathbb Z_p^*$ by
 a subgroup whose order is a prime divisor of $p-1$ or  by other
 remarkable structures as the elliptic curves group.

{\bf Received: Month xx, 200x}


\begin{thebibliography}{99}

\bibitem{1}
D.~Bleichenbacher,  {\em Generating ElGamal signatures without
knowing the secret key}, In Advances in Cryptology, Eurocrypt'96,
LNCS 1070, Springer-Verlag, (1996), 10 - 18.

\bibitem{2}
W.~Diffie and M.~E.~Hellman, {\em New directions in cryptography},
IEEE Transactions on Information Theory, vol. IT-22, (1976), 644 -
654.

\bibitem{3}
T.~ElGamal, {\em A public key cryptosystem and a signature scheme
based on discrete logarithm problem}, IEEE Trans. Info. Theory,
IT-31, (1985), 469~-~472.

\bibitem{4}
P.~Horster, M.~Michels, H.~Petersen, {\em Generalized ElGamal
signature schemes for one message block}, Technical Report,
TR-94-3, 1994.

\bibitem{5}
E.~S.~Ismail, N.~M.~F.~Tahat and R.~R.~Ahmad,  {\em A new digital
signature scheme based on factoring and discrete logarithms }, J.
of Mathematics and Statistics ({\bf 4}): (2008), 222 - 225.

\bibitem{6}
O.~Khadir, L.~Szalay, {\em Experimental results on probable
primality}, Acta Univ. Sapientiae, Math. {\bf 1}, no. 2, (2009), 161 - 168.\\
Available at http://www.emis.de/journals/AUSM/C1-2/math2-6.pdf


\bibitem{7}
A.~J.~Menezes, P.~C.~van Oorschot and S.~A.~Vanstone, {\em
Handbook of applied cryptography}, CRC Press, Boca Raton, Florida,
1997.\\
Available at http://www.cacr.math.uwaterloo.ca/hac/

\bibitem{8}
National institute of standard and technology (NIST). FIPS
Publication 186, DSA, Department of commerce, 1994.\\
http://www.itl.nist.gov/fipspubs/fip186.htm

\bibitem{9}
P.~Q.~Nguyen and I.~E.~Shparlinski, {\em The insecurity of the
digital signature algorithm with partial known nonces}, J. of
Cryptology, Vol. {\bf 15},  (2002), 151 - 176.

\bibitem{10}
H.~Ong, C~.P~. Schnorr  and A.~Shamir,  {\em Efficient signature
schemes on polynomial equations}, In Advances in Cryptology,
Crypto'84, LNCS 196, Springer-Verlag, (1985), 37 - 46.

\bibitem{11}
D.~Pointcheval and J.~Stern,  {\em Security proof for signature
schemes}, In Advances in Cryptology, Eurocrypt'96, LNCS 1070,
Springer-Verlag, (1996), 387 - 398.

\bibitem{12}
M.~O.~Rabin, {\sl Digitalized signatures and public key functions
as intractable as factoring}, MIT/LCS/TR, Vol. 212, 1979.

\bibitem{13}
R.~Rivest, A.~Shamir and L.~Adeleman,  {\em A method for obtaining
digital signatures and public key cryptosystems}, Communication of
the ACM, Vol. no 21, (1978), 120 - 126.

\bibitem{14}
C.~P.~Schnorr,  {\em Efficient signatures generation by smart
cards }, In Advances in Cryptology, Crypto'89, LNCS 435,
Springer-Verlag, (1990), 239~-~252.

\bibitem{15}
A.~Shamir,  {\em How to prove yourself : practical solutions to
identification and signature problems}, In Advances in Cryptology,
Crypto'86, LNCS 196, Springer-Verlag, (1987), 186 - 194.

\bibitem{16}
D.~R.~Stinson, {\em Cryptography, theory and practice}, Third
Edition, Chapman \& Hall$/$CRC, 2006.















\end{thebibliography}
\end{document}